\documentclass[namedreferences,hyperref,optionalrh,solaromanenum]{spr-sola}

\usepackage{graphicx}                    
\usepackage{color}                       

\definecolor{darkgreen}{rgb}{0,0.6,0}
\newcommand{\RT}[1]{{#1}}
\newcommand{\removed}[1]{{}}

\newcommand{\dxdycz}[3]{\left( \frac{ \partial #1 }{ \partial #2 }\right)_{#3}}

\chardef\us=`\_

\newcommand{\be}[1]{\begin{equation}\label{#1}}
\newcommand{\ee}{\end{equation}}
\newcommand{\bea}[1]{\begin{eqnarray}\label{#1}}
\newcommand{\eea}{\end{eqnarray}}

\newcommand{\kB}{k_{\rm B}}
\newcommand{\lbar}{\lambda\kern-.9ex\raise.4ex\hbox{-}}
\newcommand{\gee}{\gamma_{\rm ee}}
\newcommand{\lDH}{\lambda_{\rm DH}}

\begin{document}

\begin{frontmatter}

\title{Various Modifications to Debye-H{\"u}ckel Interactions\\
       in Solar Equations of State}

%
\author[addressref={aff1},corref,email={RTrampedach@SpaceScience.org}]{\inits{R.}\fnm{Regner}~\snm{Trampedach}\orcid{0000-0003-0866-6141}}
\author[addressref={aff2},corref,email={dappen@usc.edu}]{\inits{W.}\fnm{Werner}~\snm{D{\"a}ppen}}

%

\address[id={aff1}]{Space Science Institute, 4765 Walnut Street, Boulder, CO 80301 USA}
\address[id={aff2}]{Department of Physics and Astronomy, USC, Los Angeles, CA 90089-0484, USA}

\begin{abstract}
The first order effect of Coulomb forces between the charged particles of a
plasma is the well-known Debye-H{\"u}ckel-term. It is a negative contribution to
the pressure and energy of the gas, that at high densities will overwhelm the
ideal gas contributions and make the gas implode into a black hole. Nature
obviously constrains this term, avoiding this fate, but how? We investigate
three different mechanisms and their effects on the equation of state and on
solar models, and the physical justifications for each of them.
We conclude that higher order Coulomb terms in combination with quantum
diffraction of electrons, provide the needed convergence.
\end{abstract}

%
\keywords{Atomic processes -- Equation of state -- Plasmas}

\end{frontmatter}

\section{Introduction and Motivation}
\label{sect:intro}

The equation of state (EOS) is a fundamental ingredient in any astrophysical
modeling, both needed for the thermodynamics of the constituent gas of the
object under study, but also as a foundation for opacity calculations, needed
for radiative transfer calculations through that same body.

As we now find ourselves in the age of asteroseismology
\citep[e.g.,][]{Asteroseismology,jcd:AsteroSeism,di_mauro:AsteroseismReview},
much higher demands are placed on our models. To meet these demands we must
include a much more comprehensive array of realistic physics into our modeling.
For the EOS, one such aspect is the thermodynamic impact of Coulomb
interactions. The first-order term, derived by \citep{Deb:Huck},
has been included for decades now \citep{rouse:FirstDH-EOS}, and since then
it has also been known to give rise to negative total pressures for high
densities and low temperatures.

\RT{This is because the first order term is
negative and its magnitude increase faster than the other pressure-components.
The next term in a density expansion would be positive and prevent a collapse
of the plasma by ensuring a positive pressure. It will also diverge for
slightly higher
densities and temperatures, than before. In other words the expansion is only slowly
converging, posing a real problem for high density EOS calculations, such as
are needed for low mass stars, planets, white dwarfs, and even the Sun,
given the demanding accuracy of helioseismic inversions (as illustrated in the
lower right-hand panel of Fig.\ \ref{fig:suncmp_4plot}).
Remedies for this has been sought, and there are some alternatives to the
systematic, but slowly converging expansion in density. }

We explore and illustrate some mechanisms that have been used to prevent the
negative pressure issue using the example of the MHD EOS
\citep{mhd1,mhd2,mhd3}, and its descendant the, {\it T}-MHD EOS, currently in
the final stages of development by the authors. \RT{All the EOS comparison
cases presented here, except for the OPAL EOS
{\protect\citep{rogers:OPAL-EOS,rogers:newOPAL-EOS}}
have been computed with the {\it T}-MHD code.}

\section{The Helmholtz Free Energy \RT{and} Coulomb Interactions}
\label{sect:TabMerge}

\RT{The {\it T}-MHD EOS is 
based on the thermodynamic potential of the \emph{Helmholtz free energy}, $F$.
From this potential, all other thermodynamic variables can be computed as
derivatives, e.g.,
\be{eq:Fders}
    p = -\dxdycz{F}{V}{T,\{N\}}\ ,\quad
    S = -\dxdycz{F}{T}{V,\{N\}}\ ,\quad
\ee
for the pressure, $p$, and the entropy per volume, $S$,
where $\{N\}$ denotes under the constraint of equilibrium densities of all
species of particles.
These are first-order thermodynamic derivatives. Among the second-order
thermodynamic derivatives we find the heat capacities, expansion coefficients,
and the adiabatic exponent,
\be{eq:gm1}
    \gamma_1 = \dxdycz{\ln p}{\ln\varrho}{\rm ad}\ .
\ee
For most modern EOS, including the {\it T}-MHD, both first-
and second-order derivatives are analytical, ensuring both thermodynamic
consistency, smoothness and enabling stable numerical differentiation of
second-order thermodynamic derivatives. It also ensures the quality of the
comparisons presented below.}

\RT{An ideal gas is characterized by ignoring the Coulomb interactions between free
charges in the plasma (in spite of such forces being needed for thermalization
of the gas by collisions). The effects of including Coulomb interactions
increase with the density of such charges.
To first order, this} is the Debye-H{\"u}ckel-term \citep{Deb:Huck},
\be{eq:FDH}
    F_{\rm DH} = - {\textstyle\frac{1}{3}}N_{\rm ion}\kB T\Lambda\ ,
\ee
for temperature $T$, and ion number density, $N_{\rm ion}$. The plasma coupling
parameter is,
\be{eq:Lambda}
    \Lambda = \frac{V}{4\pi N_{\rm ion}}\lDH^{-3}\ ,
\ee
where the Debye screening-length, including electron degeneracy via
$\theta_{\rm e}$, is
\be{eq:lDH}
    \lDH^{-2} = \frac{4\pi e^2}{\kB T}\left[
    N_{\rm e}\theta_{\rm e}(\eta_{\rm e})
    + \sum_{\alpha\ne{\rm e}} Z_\alpha^2 n_\alpha\right]\ .
\ee
\RT{The non-relativistic version of the degeneracy factor is
\be{eq:theta-NR}
    \theta_{\rm e}(\eta) = \mathcal{F}_{-1/2}(\eta)/2\mathcal{F}_{1/2}(\eta)
\ee
\citep{dewitt:e-deg-gas}, and the degeneracy parameter $\eta=\mu/\kB T$ where
$\mu$ is the chemical potential of the electrons and $\mathcal{F}_\nu(\eta)$
are the Fermi-Dirac integrals.}

Since $F_{\rm DH}$ increasingly over-estimates the Coulomb effects with
increasing density, and since it provides a negative pressure contribution,
$F_{\rm DH}$ needs to be limited by some mechanism, in order to describe a
stable, non-collapsing plasma. 

We investigate three moderating factors to apply to $F_{\rm DH}$, for this
purpose and, importantly, also to obtain a far better approximation to the
plasmas of low-mass stars. The following EOS calculations are performed
for a H-He mixture of $X=71.6$\% and $Y=28.4$\% by mass, respectively
(10:1 ratio by number).

\RT{We show comparisons between only two of the four independent thermodynamic
variables: the pressure, $p$, which is crucial for the hydrostatic support of
stars, and the adiabatic exponent, $\gamma_1$, since it can be inverted for in
helioseismic analyses \citep{sb-jcd:eos-osc,dimauro:high-l-suncmp}. Combined
with the fact that below the top 2--3\,Mm, the rest of the solar convective
envelope is exceedingly close to adiabatic, and its stratification therefore
almost entirely determined by the EOS and the composition, makes $\gamma_1$
in this region a powerful probe of the EOS.}

The {\it T}-MHD EOS, used for these calculations, is a descendant of the MHD EOS
with many updates to the employed physics. The choices of what physics to
use, is controlled by flags, including how $F_{\rm DH}$ should be
modified. This flexible approach enables a study like the present, exploring
and comparing different choices for isolated parts of the EOS. It will be
presented in a future series of papers.

\subsection{The $\tau$-correction}
\label{sect:tau}

The so-called $\tau$-correction was first introduced by
\citet[][GHR]{graboske:Fmin+tau}, as a means of truncating the diverging
\citet{Deb:Huck}-term, $F_{\rm DH}$\removed{, which will lead to a negative total
pressure at high density, if left unrestrained}.
This factor was subsequently adopted in the MHD EOS by \citet{mhd1}, and also
the Chem EOS by \citet{kilcrease:ChemEOS-F4}.
A discussion of this term can also be found in \citet{rt:EOS-comp}.

GHR argued that there would be a minimum distance of
approach between ions and electrons of
\be{eq:rmin}
    r_{\rm min} = e^2\langle Z\rangle
   \left[\kB T\frac{\mathcal{F}_{3/2}(\eta_{\rm e})}{\mathcal{F}_{1/2}(\eta_{\rm e})}\right]^{-1}\ ,
\ee
being the distance at which the kinetic energy of an average electron (the
square parenthesis) equals the potential energy around an ion of average
charge, $e^2\langle Z\rangle$. The use of Fermi integrals, $\mathcal{F}_\nu$, allows for
arbitrary degeneracy of the electrons. While this \RT{$r_{\min}$} could make sense for
interactions between ions, it is unclear how this would provide a closest
approach between oppositely charged particles.

Proceeding with this $r_{\rm min}$, GHR used the formulation developed as part
of the original \citet{Deb:Huck} theory of electrolytes, to account for hard
sphere behavior of constituent molecules, which results in a factor
\be{eq:taux}
    \tau(x) = 3[\ln(1+x)-x+x^2/2]x^{-3}\ ,\qquad
    x = r_{\rm min}/\lDH\ ,
\ee
Ignoring the various degeneracy factors, and assuming electron density,
$N_{\rm e}$ to be proportional to density, $\varrho$, we see that
$x\propto \varrho^{-1/2}T^{-3/2}$.

The form of the $\tau$-factor arise from the so-called \emph{recharging}
procedure \citep{Deb:Huck}, which evaluates the free energy of a potential 
as an integral of that potential over the charge, from zero to full charge.
The procedure is carried out for the screened point-charge potential and the
screened, finite volume charge potential, and $\tau(x)$ is the ratio between
the two. This recharging process is only valid, however, if $r_{\rm min}$
does \emph{not} depend on charge, which Eq.\ (\ref{eq:rmin}) plainly does.

\begin{figure} 
\leftline{\hspace{-1.10em}{\bf a)}}\vspace{-0.7\baselineskip}
\centerline{\includegraphics[width=0.5\textwidth,clip=false]{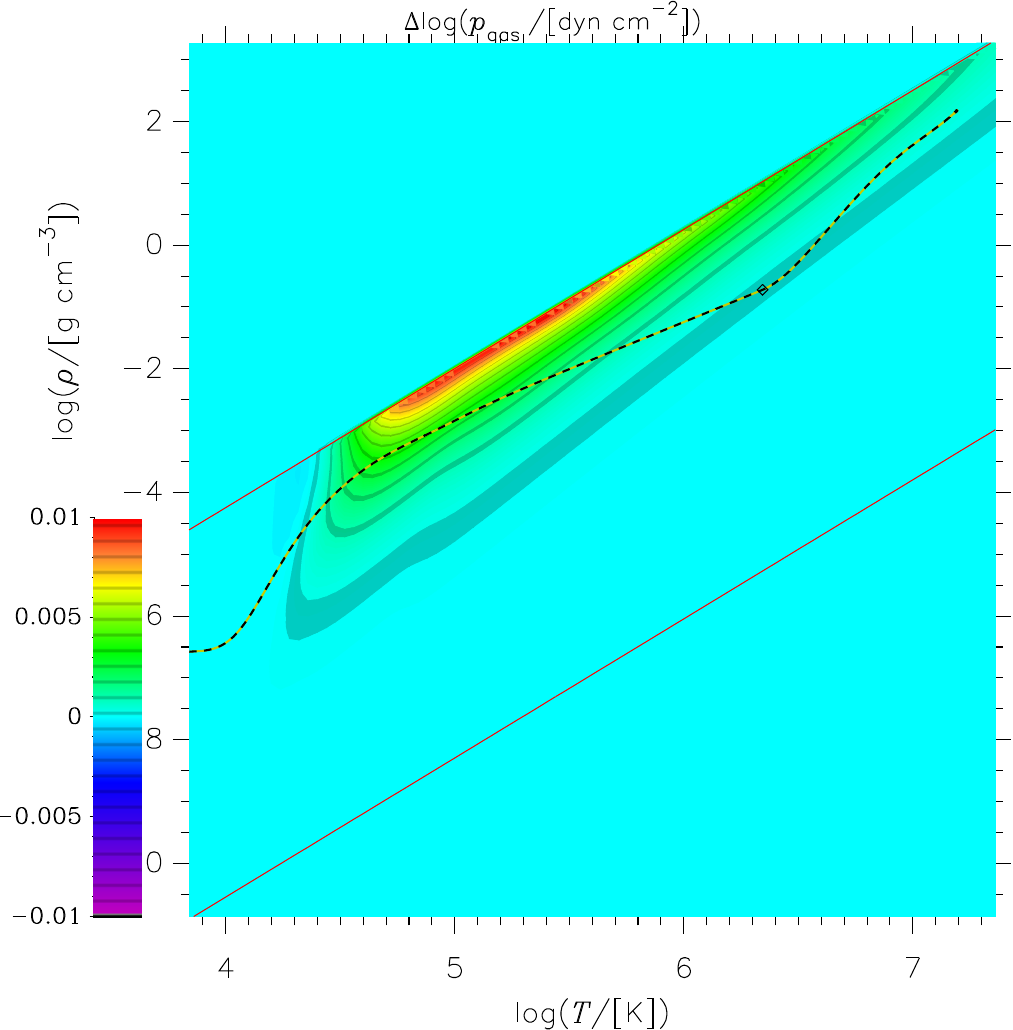}
            \includegraphics[width=0.5\textwidth,clip=false]{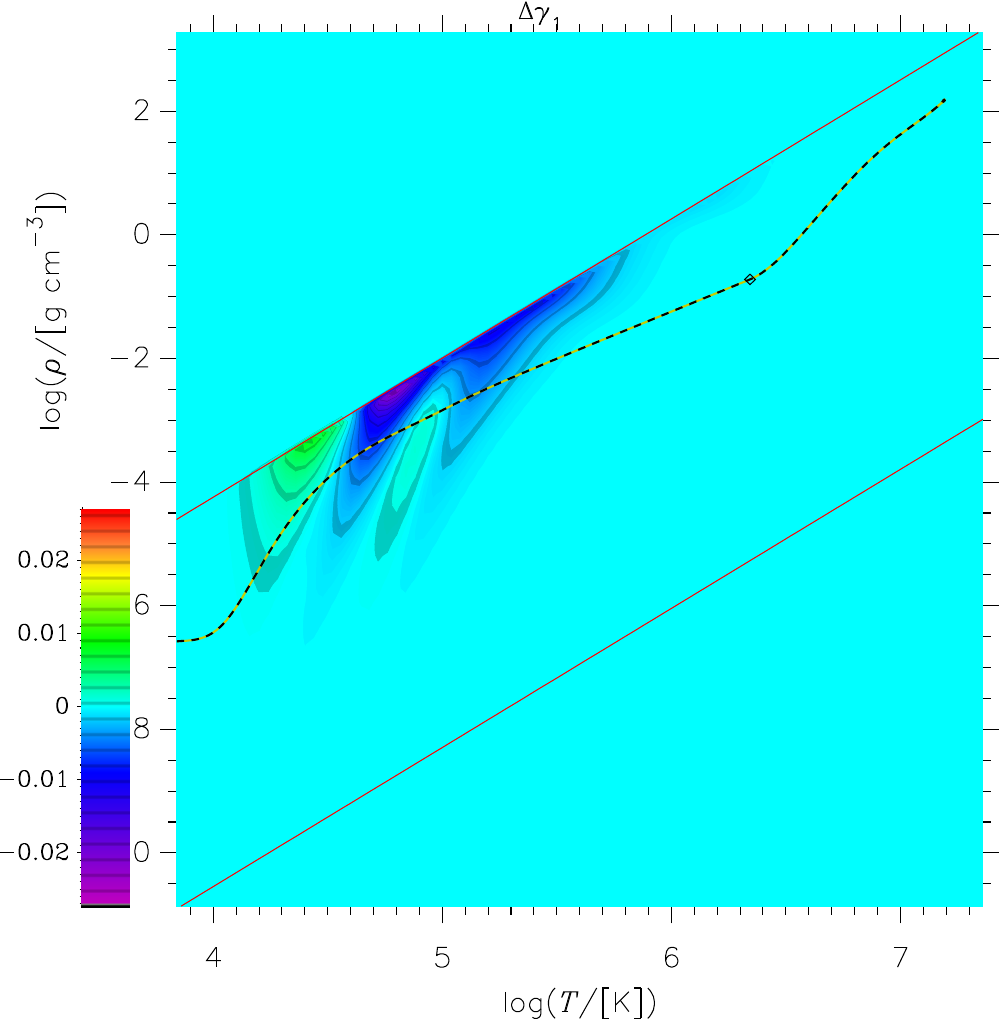}}
\leftline{\hspace{-1.20em}{\bf b)}}\vspace{-0.7\baselineskip}
\centerline{\includegraphics[width=0.5\textwidth,clip=false]{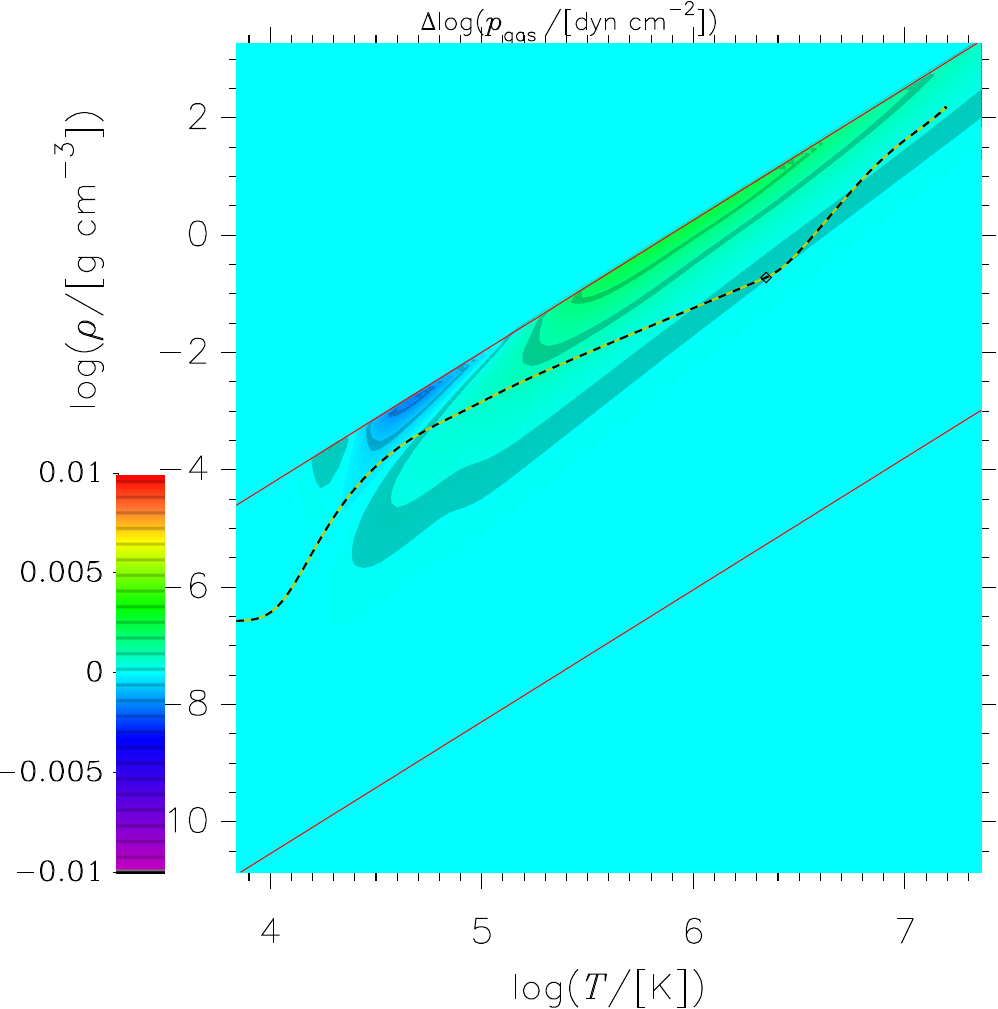}
            \includegraphics[width=0.5\textwidth,clip=false]{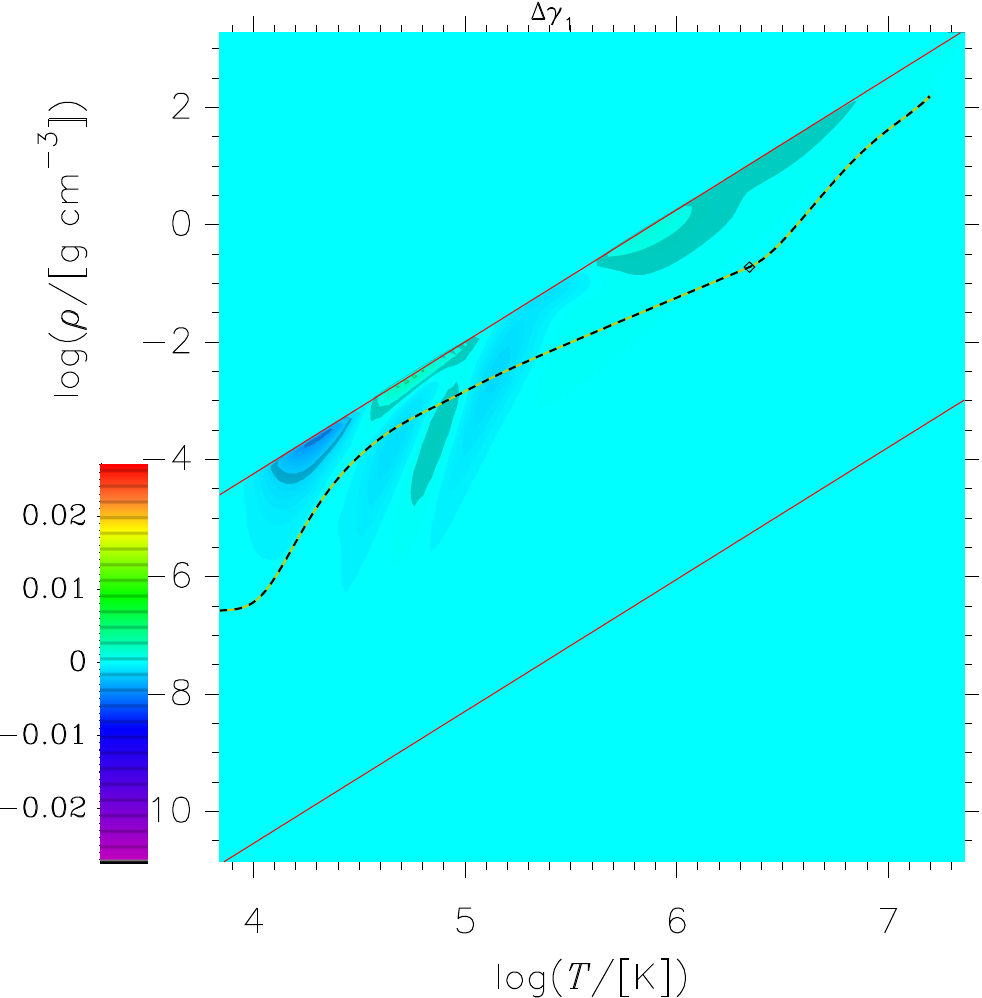}}
\leftline{\hspace{-1.05em}{\bf c)}}\vspace{-0.7\baselineskip}
\centerline{\includegraphics[width=0.5\textwidth,clip=false]{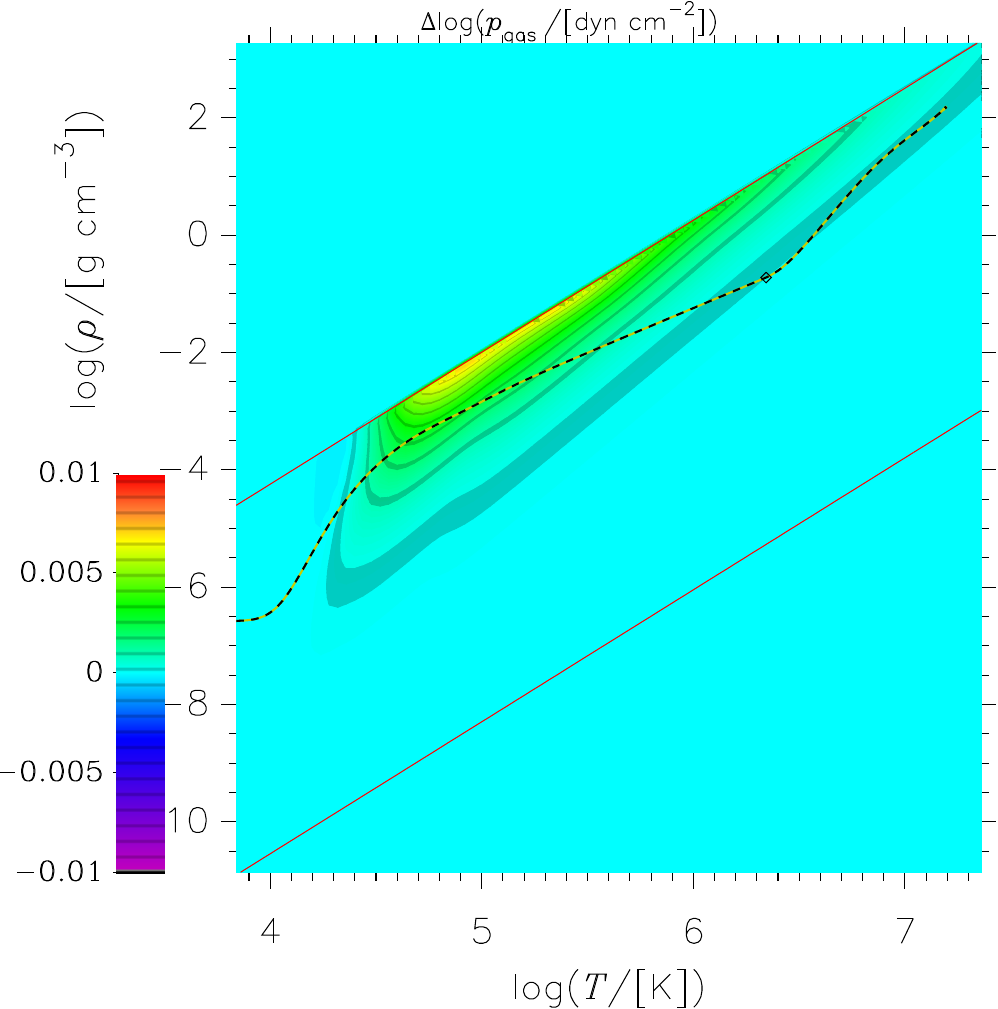}
            \includegraphics[width=0.5\textwidth,clip=false]{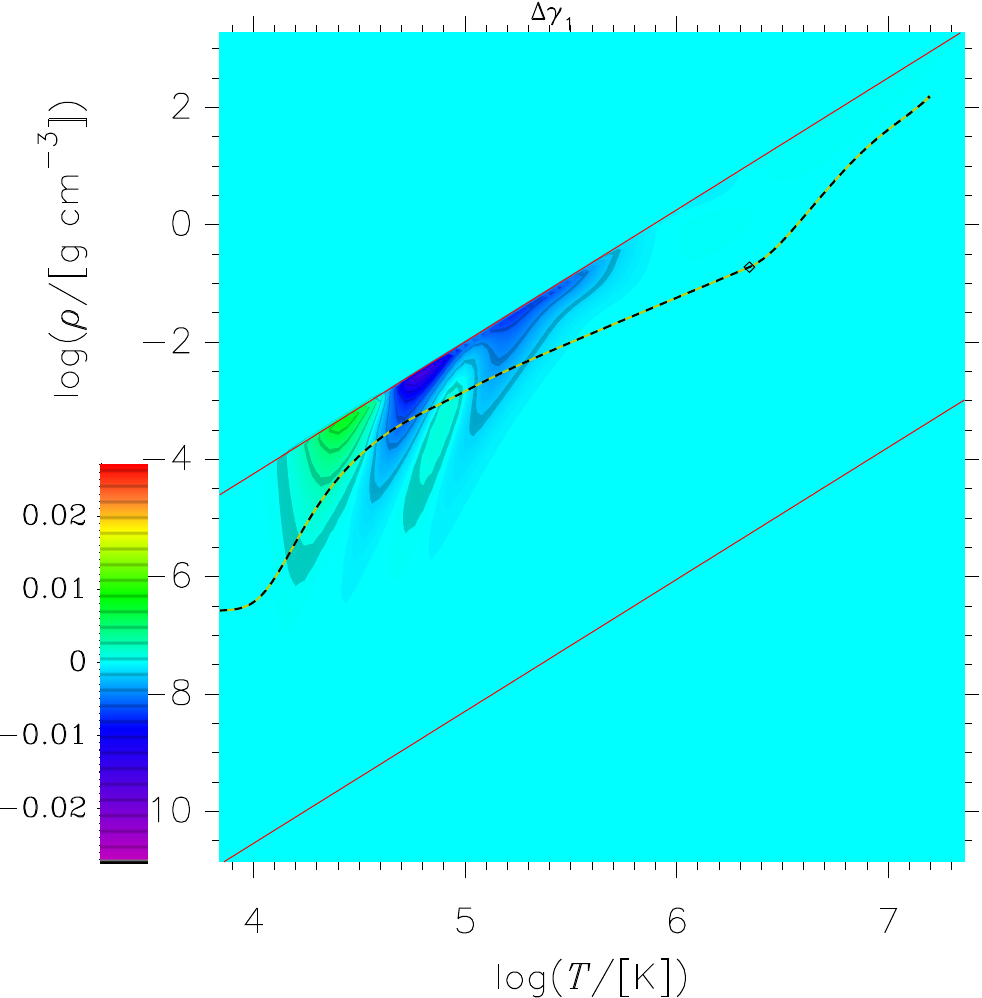}}
\caption{
The shading shows the differences between different modifications of
the DH-term in the three rows, in the sense 
\RT{
$F_{\rm DH}\!\times\!\left[\{1,\tau(x),q(\gamma_{\rm ee})\}-g(\Lambda)q(\gamma_{\rm ee})\right]$,
comparing three cases against the $g(\Lambda)q(\gamma_{\rm ee})$-case used for
{\it T}-MHD}. Turquoise shows the zero-level in all panels, but there are
separate scales for the left and right columns, as shown with the color-bars
for each panel. \RT{{\bf a)} the bare $F_{\rm DH}$, {\bf b)} $\tau(x)$ (see
Sect.\ \ref{sect:tau}), {\bf c)} quantum diffraction, only (see
Sect.\ \ref{sect:qdiff}), effectively showing the consequences of higher-order
Coulomb terms, $g(\Lambda)$ (see Sect.\ \ref{sect:Coul}).} The left column shows
gas pressure differences, and the right column shows $\gamma_1$ differences. The
dashed line shows the stratification of solar Model\,S by
\citet{GONG-Sci:sol-mod}. The inflection around $\log T=6.4$, mark\RT{ed with a
small $\diamond$-symbol, is} the bottom of the convective envelope, where the
temperature gradient changes from adiabatic to radiative. Diagonal lines show
the extent of the tables.}
\label{fig:tausTabEffect}
\end{figure}

The effect of using the $\tau(x)$-factor on $F_{\rm DH}$ is shown in
Fig.\ \ref{fig:tausTabEffect}{\bf a)}. As with all these $F_{\rm DH}$
modifications $\tau(x)$ moderates the negative DH pressure, increasing the
total pressure. Of the three modifications we explore here, this one has the
largest amplitude, in both pressure, $p_{\rm gas}$, and adiabatic exponent,
$\gamma_1$.
The iso-contours of this $\tau(x)$-effect approximately follow
lines with $\varrho\propto T^{2.6}$.

\subsection{Quantum Diffraction}
\label{sect:qdiff}

This is a manifestation of \citet{heisenberg:uncert}'s uncertainty principle,
effectively turning the classical point charge into an extended distribution
of particle with finite charge density, $\rho$. This has the profound effect
that two charged, quantum particles can get arbitrarily close, and instead of
diverging, the electrostatic potential between them converges to zero.
The result of this is a smooth, short-range truncation of the coulomb
interactions, as
investigated by, e.g., \citet{riemann:1comp-plasma}. An analytical fit\RT{,
$q(\gamma_{\rm ee})$,} to
their results were adopted in the {\it T}-MHD EOS \citep{trampedach:T-MHD}.
The argument to this factor is the de\,Broglie wavelength between two
electrons
\bea{eq:deBroglie}
    \nonumber
    \lbar_{\rm ee}^2 = \frac{\hbar^2}{\kB T}\frac{\Theta_{\rm e}}{m_{\rm e}}\ ,
\eea
in units of the Debye-length, Eq.\ (\ref{eq:lDH}), $\gee=\lbar_{\rm ee}/\lDH$.
The factor, $\Theta_{\rm e}$, corrects for electron degeneracy. Ignoring\RT{,
for the moment,}
$\theta_{\rm e}$ and $\Theta_{\rm e}$ we find that
$\gee\propto\sqrt{\varrho}/T$.

\begin{figure} 
\centerline{\includegraphics[width=\textwidth,clip=false]{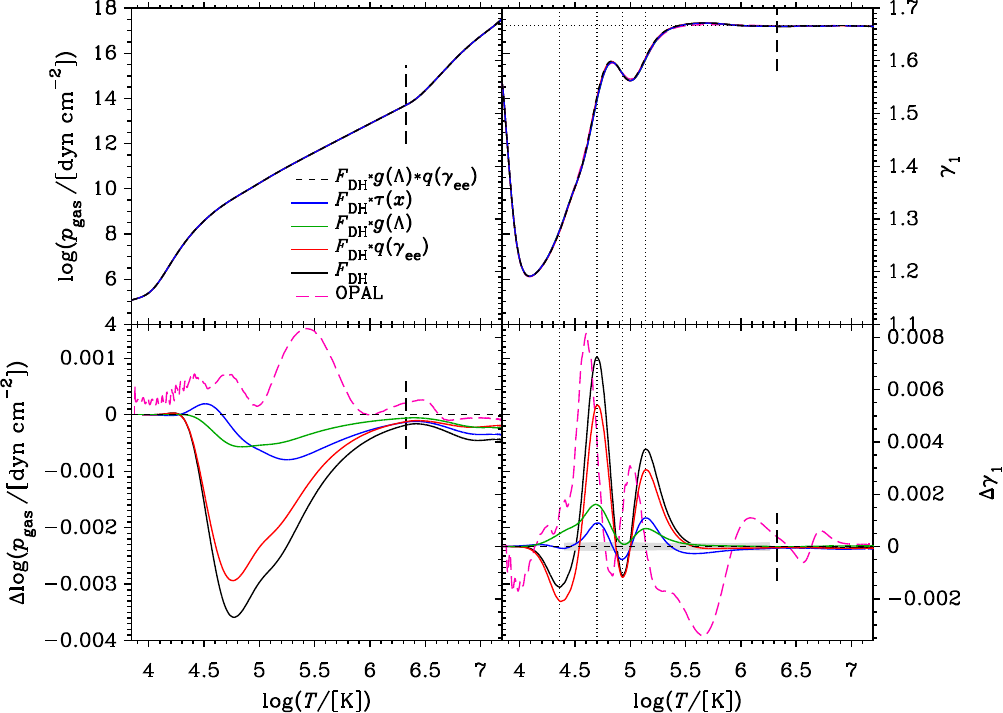}}
\caption{The differences shown in Fig.\ \ref{fig:tausTabEffect}, but for the
solar stratification, also shown in Fig.\ \ref{fig:tausTabEffect}.
The various cases\RT{, $X$ as list in the top-left panel,} are
{\protect\removed{\emph{not} over-}}\RT{all} plotted in the top two plots of
absolute $\log p_{\rm g}$ and $\gamma_1$, {\protect\removed{since they are too
close to distinguish}}\RT{but they can only be distinguished at high
magnification}. The dotted horizontal line in the top-right plot shows the
non-ionizing perfect gas value of $\gamma_1=5/3$.
The bottom two plots shows the effects of applying the $\tau(x)$, $g(\Lambda)$,
quantum diffraction and the combined Coulomb and quantum diffraction factors
to the {\protect\citet{Deb:Huck}}-term, in the sense \RT{(case $X$ minus {\it T}-MHD).
The long-dashed magenta lines show the OPAL EOS for comparison
\citep{rogers:newOPAL-EOS}. The high-frequency noise at low $T$ is interpolation
errors from the OPAL routines}.
The vertical dotted lines in the right-hand-side plots show the location of
extrema in $\Delta\gamma_1$. \RT{The vertical, long-dashed line-segments show the
location of the bottom of the convective envelope.}
The gray band around the zero-line of the
$\Delta\gamma_1$ plot shows the error-bars from a helioseismic inversion of
$\gamma_1$, by {\protect\citet{dimauro:high-l-suncmp}}, highlighting how observationally
significant the EOS differences are.}
\label{fig:suncmp_4plot}
\end{figure}

This factor goes from 1 to 0, asymptotically going to zero as $\gee^{-1/2}$ for
large $\gee$, but applies only to the electron-electron part of $F_{\rm DH}$\RT{,
which we indicate with an asterisk, $*$, instead of an ordinary multiplication}.
The effect of electron-electron diffraction is shown for the tables in Fig.\
\ref{fig:tausTabEffect} \RT{as the difference between rows a-c} and iso-contours
are found to approximately follow
$\varrho\propto T^{2.2}$, which is shallower than for $\tau(x)$, Sect.\
\ref{sect:tau}. The effect on the solar case is shown in Fig.\
\ref{fig:suncmp_4plot}, \RT{as the difference between the black and red solid
curves,} and is only about a third of the other two cases\RT{, $\tau(x)$:
difference between solid black and blue, $g(\Lambda)$: solid black and green},
both for the pressure and the adiabatic exponent.

For the interaction between identical particles, there is an additional quantum
mechanical term describing the exchange effect, $F_{\rm ex}$, which is a
consequence of \citet{pauli:excl}'s exclusion principle; If two identical
fermions meet they will either form a symmetric wave-function (when they form
an anti-symmetric singlet spin state) or an anti-symmetric wave-function (when
they form a symmetric triplet spin state). The latter case will prevent any
overlap between the two particles, thereby further reducing the interaction
energy between them. This is not formulated as a factor on the
Debye-H{\"u}ckel term, and \removed{we do not include it}\RT{although it is
included in all the {\it T}-MHD calculations shown here, it is not addressed}
in the present analysis.

\subsection{Higher-order Coulomb interactions}
\label{sect:Coul}

The long-range aspect of Coulomb forces greatly complicates the treatment of
Coulomb interacting systems. Only crystalline structures can be handled
analytically in closed form, due to their spatially
periodic structure. For gases and the high-density transition to a liquid, there
is no such simplifying mechanism. That leaves us with two tools for
investigating these systems. Analytical expansions in density (or equivalent
quantity), or simulations (e.g., Monte Carlo or Molecular Dynamics). Analytical
expansions can be derived rigorously and include all the physics to a given
order, but the complexity rises exponentially with order, quickly making this
approach prohibitively labor intensive. So far expansions up to order
$\varrho)^{5/2}$ have been accomplished
\citep{dewitt:EOS-ne5,alastuey:Fermi-BoseStats} and are available to astronomers
\citep{rogers:OPAL-EOS,rogers:newOPAL-EOS}. Despite this
effort, an order 5/2 in density is unlikely to suffice for low mass red dwarfs
and brown dwarfs, and will then diverge rapidly. The simulation alternative can
be applied to any conditions needed, but are instead limited by the
computational cost to run them with enough particles to make statistically
accurate results, and avoid boundary effects. Many such simulations are usually
needed in order to populate parameter space to strongly constrain fits to the
effect under investigation. We have chosen the simulation approach in order to
obtain an EOS with as wide a region of applicability/validity, as possible. We
employ the Monte Carlo simulations by \citet{slattery:OCP-MC} and
\citet{stringfellow:OCP-MC}, to which we have fitted our own expression for the
Coulomb factor, $g(\Lambda)$, also constrained by the low density
\citep{abe:cluster-expan} cluster-expansion. $\Lambda$ is the plasma coupling
parameter. At small $\Lambda$, $g\propto -\Lambda$ and for $\Lambda$
approaching the crystallization limit of $\Lambda_{\rm cr}=4113$, the Monte
Carlo simulations give $g\propto \Lambda^{-0.31540}$.

The effect of \RT{omitting} our implementation of $g(\Lambda)$, is shown in
Fig.\ \ref{fig:tausTabEffect}{\bf c)}. \RT{The effect of omitting the combined
effects of quantum diffraction and higher order Coulomb terms,
$q(\gamma_{\rm ee})*g(\Lambda)=${\it T}-MHD, is shown in
Fig.\ \ref{fig:tausTabEffect}{\bf a)}. The difference, $\tau(x)-
q(\gamma_{\rm ee})*g(\Lambda)$ of Fig.\ \ref{fig:tausTabEffect}{\bf b)}, shows
that $\tau(x)$ generally gives higher pressure than {\it T}-MHD, with increasing
density, and}
\removed{both the pressure and $\gamma_1$
difference are a bit smaller than with $\tau(x)$, but they also both extend to
lower densities. In other words,} the transition is smoother in density
with $g(\Lambda)$ than with $\tau(x)$.
Iso-contours of the effect of $g(\Lambda)$ approximately follows
$\varrho\propto T^{2.6}$ just as with $\tau(x)$, and is steeper than for
quantum diffraction, Sect.\ \ref{sect:qdiff}.

\section{Discussion and Conclusion}
\label{sect:conclusion}

We have investigated three different modifications to the first-order
Coulomb term of the Helmholtz free energy. The first one, the $\tau(x)$
factor, purports to model the effects of a limit to how close ions and
electrons can approach each other at a given set of plasma conditions.
Both the argument and the derivation are found to be flawed, however.

The second modification is the quantum diffraction, which describes the
quantum-mechanical analogy to a closest approach, provided by Heisenberg's
uncertainty relation, turning classical point-particles into charge
distributions of finite size. As, e.g., electrons overlap, they feel less
and less charge interior to their location, due to the everywhere finite
charge density. The effect of this is only about a third of the effect from
$\tau(x)$, and has a different shape. It is also a physical effect that
needs to be included in the EOS if aiming for high-fidelity models of
low-mass stellar objects, or comparing against helioseismology.

The last modification is that due to the higher-order terms in the
Coulomb interactions, in this case found from Monte Carlo simulations. 
The effect of this is curiously similar to that of $\tau(x)$, although
less steep in density, and of slightly lower amplitude. This factor is
also a physical effect and it needs to be combined with quantum diffraction
as shown in Fig.\ \ref{fig:suncmp_4plot} for the solar case. This combination
has a slightly \emph{larger} amplitude in a solar model, than does the 
$\tau(x)$ factor. 

The effect on the adiabatic exponent, $\gamma_1$, is very similar in shape
for all three cases, as seen in Fig.\ \ref{fig:suncmp_4plot}, except quantum
diffraction which only \emph{reduces} $\gamma_1$. From the vertical dotted
lines in the $\gamma_1$ panels of Fig.\ \ref{fig:suncmp_4plot}, it is evident that
the differences do not constitute a simple deepening of the ionization dips
of $\gamma_1$, but is rather a tilting of the dips, to have lower
'background' on the cool side and very slightly higher 'background' on the
warm side, as well as a shift of the ionization dips towards higher $T$.
These changes to $\gamma_1$ will affect helioseismic abundance determinations
which exploit these ionization dips to determine helium
\citep{dappen:SunSeismHe,antia:sunHE+EOS,basu-antia:Y,richard:sunHe,vorontsov:SunSeismHe,houdek:Solar+StellarY} and metal abundances
\citep{lin:seismEOS,houdek:SolarZ+age,vorontsov:SeismicSolarZ,buldgen:HelioseismicAbunds}
in the close to adiabatic part of the convective envelope. This is the only
method available for a solar He determination, and an important test of
spectral abundance analysis for metals.

\citet{rt:EOS-comp} compared the OPAL and MHD EOS and found that the
$\tau(x)$ factor, used in the MHD EOS, was \removed{larger}\RT{reducing
$F_{\rm DH}$ by more} than the higher-order
effects in the OPAL EOS, which are based on analytical expansions. Because
the OPAL EOS was found to be in better general agreement with helioseismology,
\citet{rt:EOS-comp} assumed the $\tau(x)$ factor was overestimating the
modification of $F_{\rm DH}$. Based on our present results \RT{(lower-left
panel of Fig.\ \ref{fig:suncmp_4plot}}, however, it is
possible that the MHD EOS\RT{, with the $\tau$-factor,} had the better
implementation of Coulomb
interactions\RT{, as the differences with respect to the {\it T}-MHD case of
$g(\Lambda) * q(\gamma_{\rm ee})$, is quite small. Such a conclusion
assumes that {\it T}-MHD is closer to reality than OPAL, and that the many
other differences between the two EOS have smaller effects.}
This is a possibility that warrants further investigation.

%
\begin{acks}
    We thank the anonymous reviewer for constructive criticism that has
    greatly improved this paper.
    We would like to thank Ga{\"e}l Buldgen and the Sierre, CH, 2023 workshop on ``The Future of Solar Modelling'', for providing the venue and forum for the discussions that hatched the project presented here.
    This research has made extensive use of NASA's Astrophysics Data System.
\end{acks}

%
%
\begin{fundinginformation}
    RT's work was funded by NASA grants 80NSSC20K0543 and 80NSSC22K0829.
\end{fundinginformation}
%
%
%
%

%
\bibliographystyle{spr-mp-sola}
\bibliography{T-MHDtaus.bib}
%
%

\end{document}